\documentclass[aps,prl,reprint,superscriptaddress,numbers,sort&compress]{revtex4-2}
\usepackage{blindtext}
\RequirePackage[T1]{fontenc}
\RequirePackage{fix-cm}

\usepackage[american]{babel}

\RequirePackage{graphicx}
\RequirePackage{mathptmx}
\usepackage{comment}
\usepackage{braket}
\usepackage{amsmath}
\usepackage{csquotes}

\DeclareMathAlphabet{\mathcal}{OMS}{cmsy}{m}{n}

\usepackage{footmisc}
\RequirePackage{natbib}

\RequirePackage{siunitx}
\RequirePackage{amsmath}
\usepackage{caption}
\usepackage{upgreek}
\usepackage{subfig}
\usepackage{caption}
\usepackage{float}
\usepackage{ragged2e}
\usepackage{bbold}

\usepackage{bm}

\graphicspath{{./Images/}}

\usepackage{url}
\RequirePackage[colorlinks,citecolor=blue,urlcolor=blue,linkcolor=blue]{hyperref}

\begin{document}

\raggedbottom

\title{ Laboratory constraints on peV-scale mass splitting between ordinary and sterile neutron states
}

\author{N.~J.~Ayres}
\affiliation{Institute for Particle Physics and Astrophysics, ETH Zürich, 8093 Zürich, Switzerland}

\author{Z.~Berezhiani}
\affiliation{INFN, Laboratori Nazionali del Gran Sasso, Assergi, 67100 L’Aquila, Italy}

\author{G.~Bison}
\affiliation{Laboratory for Particle Physics, PSI Center for Neutron and Muon Sciences, Paul Scherrer Institute (PSI), 5232 Villigen, Switzerland}

\author{K.~Bodek}
\affiliation{Marian Smoluchowski Institute of Physics, Jagiellonian University, 30-348 Cracow, Poland}

\author{V.~Bondar}
\affiliation{Institute for Particle Physics and Astrophysics, ETH Zürich, 8093 Zürich, Switzerland} 

\author{P.-J. Chiu}
\altaffiliation[Present address: ]{Department of Physics, National Taiwan University, 106319 Taipei, Taiwan}
\affiliation{Institute for Particle Physics and Astrophysics, ETH Zürich, 8093 Zürich, Switzerland}
\affiliation{Laboratory for Particle Physics, PSI Center for Neutron and Muon Sciences, Paul Scherrer Institute (PSI), 5232 Villigen, Switzerland}

\author{M.~Daum}
\affiliation{Laboratory for Particle Physics, PSI Center for Neutron and Muon Sciences, Paul Scherrer Institute (PSI), 5232 Villigen, Switzerland}

\author{C.~B.~Doorenbos}
\affiliation{Institute for Particle Physics and Astrophysics, ETH Zürich, 8093 Zürich, Switzerland}
\affiliation{Laboratory for Particle Physics, PSI Center for Neutron and Muon Sciences, Paul Scherrer Institute (PSI), 5232 Villigen, Switzerland}

\author{S.~Emmenegger}
\affiliation{Institute for Particle Physics and Astrophysics, ETH Zürich, 8093 Zürich, Switzerland}

\author{K.~Kirch}
\affiliation{Institute for Particle Physics and Astrophysics, ETH Zürich, 8093 Zürich, Switzerland}
\affiliation{Laboratory for Particle Physics, PSI Center for Neutron and Muon Sciences, Paul Scherrer Institute (PSI), 5232 Villigen, Switzerland}

\author{V.~Kletzl}
\altaffiliation[Present address: ]{Marietta-Blau-Institute for Particle Physics, Austrian Academy of Sciences, 1010 Vienna, Austria}
\affiliation{Institute for Particle Physics and Astrophysics, ETH Zürich, 8093 Zürich, Switzerland}
\affiliation{Laboratory for Particle Physics, PSI Center for Neutron and Muon Sciences, Paul Scherrer Institute (PSI), 5232 Villigen, Switzerland}

\author{J.~Krempel}
\affiliation{Institute for Particle Physics and Astrophysics, ETH Zürich, 8093 Zürich, Switzerland}

\author{B.~Lauss}
\email[Corresponding author: ]{bernhard.lauss@psi.ch}
\affiliation{Laboratory for Particle Physics, PSI Center for Neutron and Muon Sciences, Paul Scherrer Institute (PSI), 5232 Villigen, Switzerland}

\author{D.~Pais}
\affiliation{Institute for Particle Physics and Astrophysics, ETH Zürich, 8093 Zürich, Switzerland} 
\affiliation{Laboratory for Particle Physics, PSI Center for Neutron and Muon Sciences, Paul Scherrer Institute (PSI), 5232 Villigen, Switzerland}

\author{I.~Rienäcker}
\affiliation{Laboratory for Particle Physics, PSI Center for Neutron and Muon Sciences, Paul Scherrer Institute (PSI), 5232 Villigen, Switzerland}

\author{D.~Ries}
\affiliation{Laboratory for Particle Physics, PSI Center for Neutron and Muon Sciences, Paul Scherrer Institute (PSI), 5232 Villigen, Switzerland}

\author{D.~Rozpędzik}
\affiliation{Marian Smoluchowski Institute of Physics, Jagiellonian University, 30-348 Cracow, Poland}

\author{P.~Schmidt-Wellenburg}
\affiliation{Laboratory for Particle Physics, PSI Center for Neutron and Muon Sciences, Paul Scherrer Institute (PSI), 5232 Villigen, Switzerland}

\author{K.~S.~Tanaka}
\altaffiliation[Present address: ]{Waseda University, 3-4-1 Ookubo, Shinjuku-ku, Tokyo 169-8555, Japan}
\affiliation{Laboratory for Particle Physics, PSI Center for Neutron and Muon Sciences, Paul Scherrer Institute (PSI), 5232 Villigen, Switzerland}

\author{J.~Zejma}
\affiliation{Marian Smoluchowski Institute of Physics, Jagiellonian University, 30-348 Cracow, Poland}

\author{N.~Ziehl}
\email[Corresponding author: ]{ziehln@phys.ethz.ch}
\affiliation{Institute for Particle Physics and Astrophysics, ETH Zürich, 8093 Zürich, Switzerland}

\author{G.~Zsigmond}
\email[Corresponding author: ]{geza.zsigmond@psi.ch}
\affiliation{Laboratory for Particle Physics, PSI Center for Neutron and Muon Sciences, Paul Scherrer Institute (PSI), 5232 Villigen, Switzerland}

\begin{abstract}
Sterile states of matter, represented by a parallel \enquote{mirror} sector, may contribute to the observed dark matter in the Universe.  We investigated the parameter space of neutron $(n)$ to mirror-neutron $(n')$ oscillations, in the case where the two states are not necessarily mass-degenerate, taking into account interactions in the mirror sector. By tuning the magnitude of an applied magnetic-field in the range $\SI{5}{\micro\tesla} < B < \SI{360}{\micro\tesla}$ to corresponding resonance conditions for finite mass splitting, we derive exclusion limits for the $n-n'$ oscillation time constant reaching about 20~s over the mass-difference range 0.3 $-$ \SI{22}{\pico\electronvolt}. In parts of this parameter range, our limits exceed the model-dependent neutron-star-cooling bound, providing the first experimental constraints in this scenario that are more stringent than this astrophysical estimate.

\end{abstract}

\maketitle

Starting from the fifties of the last century, the existence of a parallel sector of \enquote{mirror} particles was proposed as an exact duplicate of the ordinary Standard Model (SM) particle sector~\cite{Lee:1956qn,Kobzarev:1966qya,Blinnikov:1983gh,Volkas2003}.
Since then, its motivation was strengthened by considerations that cold mirror matter could account for dark matter~\cite{Berezhiani:2000gw,Ignatiev:2003js,Berezhiani:2003wj,Berezhiani:2005vv}.
Along with gravity, mirror particles might interact with ordinary matter via specific feeble interactions \cite{Berezhiani:2003xm,Foot:2003eq}. 
In particular, lepton and baryon violating cross-interactions between the ordinary and mirror particles can co-generate baryon asymmetries in both sectors and explain the observable and dark matter fractions in the Universe \cite{Bento:2001rc,Berezhiani:2008zza}. The same interactions can induce mixing between neutral ordinary particles and their dark mirror partners  as e.g. neutrino
mixing with mirror neutrinos which make the latter natural candidates for sterile neutrinos \cite{Akhmedov:1992hh,Foot:1995pa,Berezhiani:1995yi,Alves:2026}. 
Interestingly, experimental bounds do not exclude a sizeable mass-mixing between the neutron and its sterile mirror partner, $\epsilon_{nn'}
+{\rm h.c.}$,  
with $n-n'$ conversion time-scale much less than the neutron lifetime, $\tau_{nn'}=\hbar /\epsilon_{nn'}\sim O(1)$ s~\cite{Berezhiani:2005hv,Berezhiani:2009ldq}.
Fast $n \to n'$ transition which violates baryon number by one unit can have  
intriguing astrophysical implications~\cite{Berezhiani:2006je,Mohapatra:2005ng,Berezhiani:2018zvs,Goldman:2019dbq,Berezhiani:2020zck,McKeen:2021jbh,Berezhiani:2021src}.
It evades immediate detection since under natural conditions it is suppressed by matter and magnetic field, or perhaps by some mass splitting between $n$ and $n'$~\cite{Berezhiani:2005hv,Berezhiani:2009ldq}. However, it can be measured  under controlled experimental conditions via anomalous $n\to n'$ losses  or via a regeneration process $n\to n' \to n$, using cold and ultracold neutrons (UCN)~\cite{Berezhiani:2005hv,Pokotilovski:2006gq,Berezhiani:2009ldq,Berezhiani:2017azg}.  

A series of searches for anomalous UCN losses \cite{Ban:2007tp,Serebrov:2007gw,Serebrov:2008her,Bodek:2009zz,Altarev:2009tg,Berezhiani:2017jkn,nEDM:2020ekj,Ayres:2026lyw} were focused on the case of exact mirror parity, when $n$ and $n'$ are exactly degenerate in mass. Some of these experiments~\cite{Serebrov:2008her,Berezhiani:2017jkn} observed significant anomalies interpreted as a signal of $n-n'$ conversion in \cite{Berezhiani:2012rq,Berezhiani:2017jkn}, but the corresponding parameter space was practically excluded by the recent PSI experiment~\cite{Ayres:2026lyw}. 

However, $n$ and $n'$ can have non-zero mass splitting~\cite{Berezhiani:2000gw}, which was invoked also to explain the neutron lifetime anomaly~\cite{WietfeldtGreene:2011,Berezhiani:2018eds,Wietfeldt:2024sym}. For example, spontaneous breaking of mirror parity, 
motivated by asymmetric post-inflation reheating between the two sectors~\cite{Berezhiani:1995am}, can induce mass splitting $m_{n'}-m_n=\delta m > O(10^{-12})$~eV~\cite{Mohapatra:2017lqw,Babu:2021mjg}. 
More generally, it can be mimicked by some environmental factors, such as hypothetical  spin-independent long-range forces differently acting on $n$ and $n'$ \cite{Berezhiani:2009ldq,Addazi:2016rgo} or even by differences in the gravitational interaction between ordinary and mirror particles in the context of modified theories of gravity~\cite{Berezhiani:2009kv,Berezhiani:2010gya}. In theories with large extra dimensions, it can be induced by some difference between ordinary and mirror branes~\cite{Sarrazin:2012jr}, or the neutrons can undergo mixing with finely spaced bulk fermions of the so-called Kaluza-Klein tower~\cite{Dvali:2023zww}.

Recent experiments \cite{Almazan:2021fvo,Stasser:2020jct,Broussard:2021eyr,Ban:2023cja,Gonzalez:2024dba,
BRENOT2026140673,Hostert:2022ntu} set \enquote{wiggly} limits on $\tau_{nn'}$, reaching up to about 10~s for some narrow ranges of $\delta m$ but are much weaker elsewhere. It should be noted that there is an astrophysical bound from the neutron star cooling, which is, however, model dependent and must be interpreted as an indicative order of magnitude estimate, $\tau_{nn'} > O(10)$~s \cite{Berezhiani:2020zck,McKeen:2021jbh,Berezhiani:2021src}. This bound is important as a benchmark because it has so far been stronger than laboratory constraints in the finite-mass-splitting scenario.

To perform experiments with increased sensitivity, a collaboration at PSI assembled a dedicated apparatus described in \cite{Ayres:2021zbh,Ingo,Nathalie}. 
In this Letter, we report on a new analysis of raw data used in~\cite{Ayres:2026lyw} as well as additional data sets covering larger mass splittings to further constrain the parameter space, namely, the 
characteristic time $\tau_{nn'}=\hbar /\epsilon_{nn'}$  as a function of $\delta m$. 
We obtain new limits which in the range $\delta m =1.5\times 10^{-12} - 1\times 10^{-11} $~eV exceed the ones obtained in previous experiments ~\cite{Almazan:2021fvo,Broussard:2021eyr,Stasser:2020jct,Ban:2023cja,Gonzalez:2024dba,
BRENOT2026140673} as well as the astrophysical bound from neutron stars. 

In a storage experiment, $n\to n'$ conversion would appear as a magnetic-field-dependent deficit in the number of detected UCNs, since sterile $n'$ would no longer be confined by the material walls of the vessel.   
The evolution of a two-state $n-n'$ system is described by the non-relativistic Hamiltonian
\begin{equation}
H = 
    \begin{pmatrix}
        E_n & \epsilon_{nn'}\\
        \epsilon_{nn'} & E_{n'}
    \end{pmatrix}.
    \label{eq:mass_splitting}
\end{equation}
Here $E_n=m_n + V$, where $V$ is the potential induced by background matter 
or a magnetic field. Analogously, $E_{n'}=m_{n'} + V'$. 
The difference of energy levels $\Delta E=E_{n'} -E_n$ can be interpreted as a generalized 
mass-splitting, $\delta m = m_{n'} - m_n + V'$, an implicit function of the difference between the masses and an unknown potential $V'$, representing possible interactions in the mirror sector. 
Solving the Schrödinger equation, the probability of $n \to n'$ transition as a function of free flight time $t$ between collisions becomes:
\begin{equation}
P^{n n'}(t) =  \frac{(2\hbar/\tau_{nn'})^2}{ (V-\delta m)^2} \sin^2 \left(\frac{V-\delta m}{2 \hbar}\,t\right).
\label{eq:prob}
\end{equation}
$n\rightarrow n'$ oscillations occur at maximum amplitude in case of degeneracy, when 
$\Delta E=0$. This can be achieved by tuning, e.g., the  magnetic field, $B$, so that the
Zeeman energy $V = \pm \mu _n B$ compensates the value of $\delta m$. 
This implies that for a given sign of the mass difference only one of the neutron spin states can match the  aforementioned degeneracy condition. When unpolarized UCNs are used, only half of the neutrons can oscillate into the sterile state. The propagation time, $t$, resets upon wall collisions in the storage volume, since the reflected neutrons remain in the vessel and continue to oscillate, while the mirror neutrons escape. 

In order to search for such transitions, our experiment employed tunable magnetic fields in the range of $\SI{5}{\micro\tesla} < B < \SI{360}{\micro\tesla}$.  
We worked on the one hand with the asymmetry, $A_{i,j}$, between neutron counts from two different $i,j$ magnetic-field settings as the experimental observable. 
On the other hand, we calculated the corresponding ratio from the loss factors that might come from $n \to  n'$ transitions, taking into account that neutron beta decay and neutron capture at wall collisions do not depend on the magnetic field settings. Equating the two gives
\begin{equation}
    A_{ij} = \frac{n_i - n_j}{n_i + n_j}  = 
    \frac{\exp(-m_s P^{n n'}_{i})- \exp(-m_s P^{n n'}_{j})}
    {\exp(-m_s P^{n n'}_{i})+ \exp(-m_s P^{n n'}_{j})}.
    \label{eq:asym}
\end{equation}
Here, $n_{i,j}$ represent neutron counts after a storage time $t_s$, when applying two magnetic fields, $B_{i,j}$. During the magnetic field scan, one of the fields is meant to induce degeneracy and the other to serve as a contrasting non-degenerate reference. $P^{n n'}_{i,j}$ is the mean probability from Eq.~\eqref{eq:prob}, for a free flight time between wall collisions,  
and $m_s$ is the mean number of collisions of UCNs detected after storage time $t_s$. 

As also discussed in~\cite{Ayres:2026lyw}, in case of a realistic inhomogeneous magnetic field, one has to consider the local increment in the oscillation probability. This can be calculated as the product of the time derivative of the probability function in Eq.~\eqref{eq:prob} and an infinitesimally short time step. The oscillation probability between two bounces, after neglecting higher order terms, becomes the sum of all infinitesimal probabilities along the path. 
The total oscillation probability can be obtained by adding the probabilities between the bounces, as expressed in Eq.~\eqref{eq:asym} using mean values $m_s P^{nn'}_{i,j}$. By performing detailed Monte Carlo (MC) simulations, one can more precisely calculate the cumulated probability during the storage and emptying periods, when the UCNs experience the magnetic field.

From the squared dependence of the probability in Eq.~\eqref{eq:prob} on $\tau_{nn'}$, in linear approximation Eq.~\eqref{eq:asym} becomes:  
\begin{equation}
    \tau_{nn'} = \frac{1}{\sqrt{2|\langle A_{ij}\rangle}|}\sqrt{|F(B_i, \delta m)  -F(B_j, \delta m)|}
\label{eq:tau}
\end{equation}
The terms $F(B_i, \delta m)$, which we hereafter denote as "resonance functions",
represent the cumulated oscillation probabilities (considered as $\ll 1$) for a unit $\tau_{nn'}=1$~s, and were calculated in MCUCN~\cite{Zsigmond:2017gmi,Bison:2019hot,
Bison:2021rdj,Ingo} simulations, considering every applied magnetic-field setting separately as a function of $\delta m$~\cite{Nathalie}.

\begin{figure}
    \includegraphics[width=\linewidth]{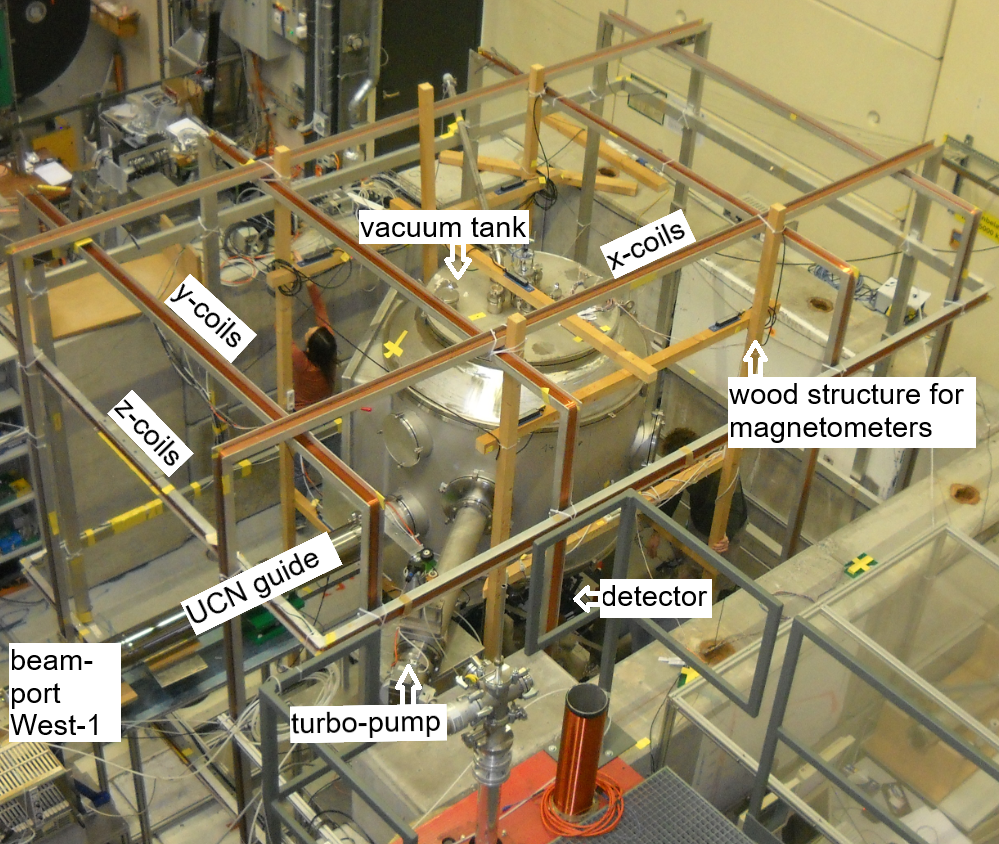}
    \includegraphics[width=\linewidth]{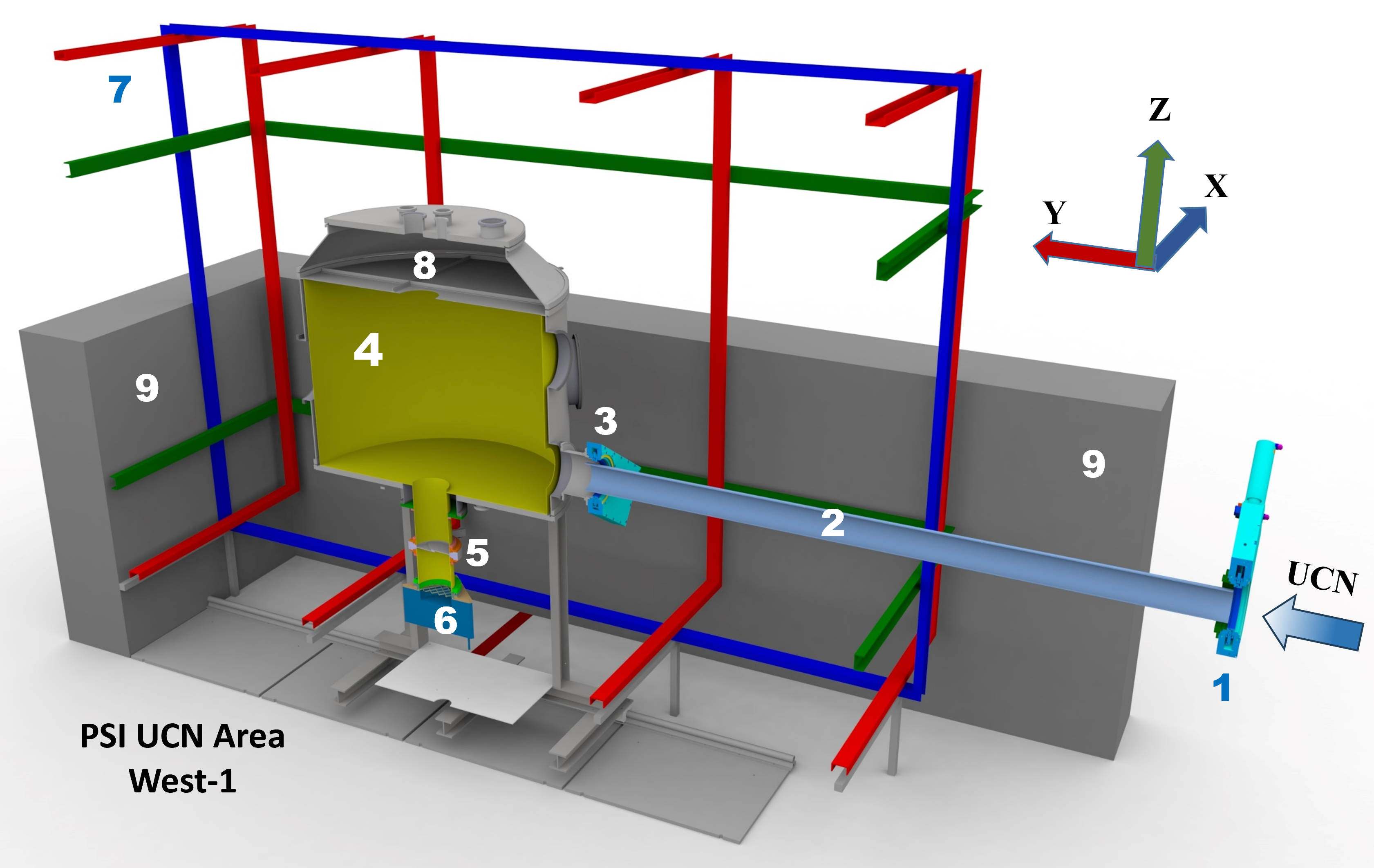}
    \captionsetup{justification=Justified}

    \caption{Experimental setup: photo and schematic. Top $-$ general view indicating the location of the main parts. Bottom $-$ cut through the schematic (CAD): (1) West-1 beamport shutter, (2) UCN guides, (3) horizontal guide shutter, (4) storage vessel, (5) butterfly shutter, (6) UCN detector, (7) magnetic field coils perpendicular to the X, Y, Z directions, (8) top lid of the storage vessel, (9) radiation shielding (adapted from \cite{Ingo}).}
    \label{fig:setup}
\end{figure}

An overview photograph and schematic of the apparatus 
(used also for experiment~\cite{Ayres:2026lyw}) 
are shown in Fig.~\ref{fig:setup}. For clarity, we briefly describe the setup installed in Area West of the PSI UCN source~\cite{Anghel:2009272,Bison:2019hot,
Bison:2021rdj,Lauss:2021rac}. During the filling period, UCNs passed through the West-1 beamport shutter (1), propagated along the Ni/Mo-coated glass guides (2), passed through the open horizontal guide shutter (3), and entered a 1~m high, 1.5 m$^3$ large storage vessel (4) with uncoated 316L stainless-steel walls of low magnetic permeability, cleaned to minimize UCN losses using a procedure described in~\cite{LGoeltl:2012}. 
The storage system was cleaned of magnetic spots (>$\SI{180}{\micro\tesla}$) by abrasion and/or using a mobile demagnetization device~\cite{WalkerHagou}, and was evacuated for the measurements. During monitoring and after storage, the neutrons were released through a fast butterfly shutter~\cite{Bison2016449} (5) and counted by a 20~cm $\times$ 20~cm GEM-based CASCADE~\cite{CASCADE2025} UCN detector (6). Eight rectangular coils (7)~\cite{Brys:2005}, around the vacuum tank of the storage vessel generated the target magnetic field during the storage and emptying periods. The top lid of the storage vessel (8) was temporarily opened to insert a device mapping the magnetic field in the storage volume. The setup was surrounded by concrete blocks for radiation protection (9).

The magnetic-field model providing input for the data analysis was based on dedicated mapping measurements inside the storage volume (using five FLC3-70 fluxgates from Stefan Mayer Instruments), supplemented by external three-axis fluxgate monitors (16 sensors, most of them Sensys FGM3D with an accuracy of $\SI{10}{\nano \tesla } \pm$ 0.1\%), as described in~\cite{Ayres:2026lyw}.
The measured response to the coil currents was characterized using a harmonic-polynomial field expansion detailed in~\cite{Abel:2018arc}. The resulting coefficients provided the position-dependent magnetic field input for the MCUCN tracking calculations, which had previously been benchmarked using this experimental setup against UCN storage measurements~\cite{Ingo}. 

\begin{figure}
    \includegraphics[width=\linewidth]{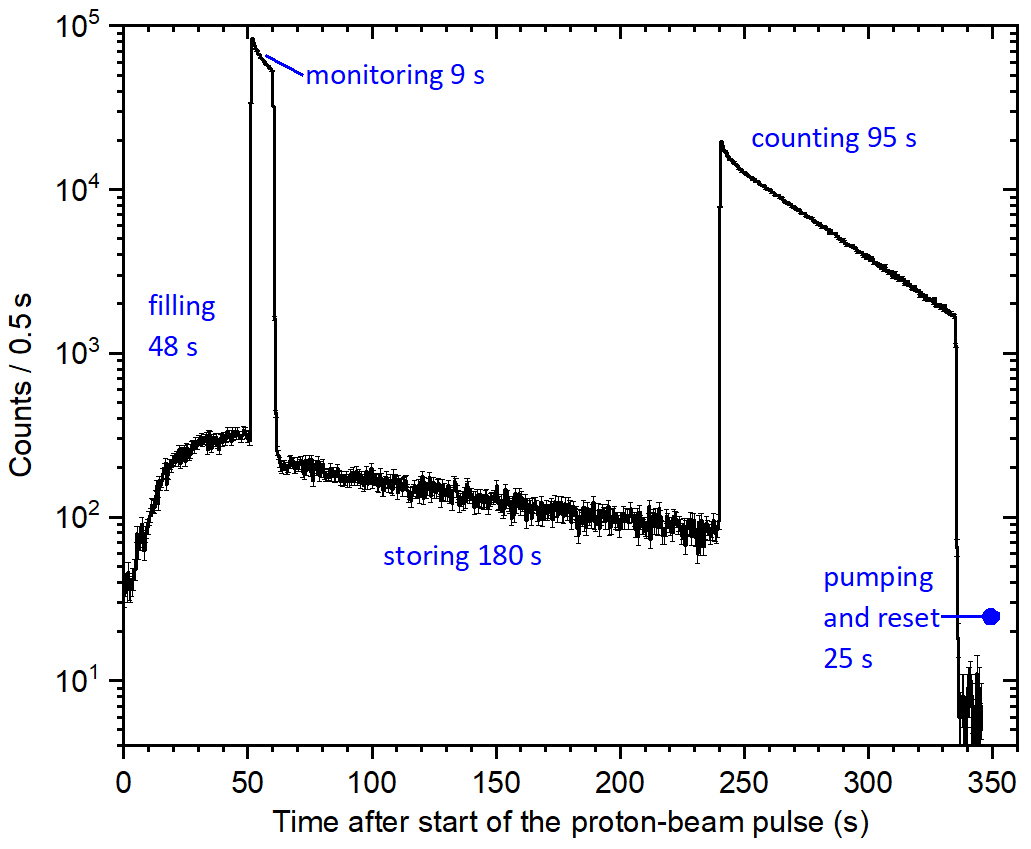}
    \captionsetup{justification=Justified}

    \caption{UCN counts versus detection time after start of the proton-pulse typically measured during one storage measurement cycle. Five 0.1~s channels were added to reduce statistical fluctuations for better visibility. The UCN counts at 'filling' and 'storing' were due to a small leakage through the butterfly shutter.}
    \label{fig:spectrum}
\end{figure}

A detailed description of the sequence of events during one measurement cycle, see Fig.~\ref{fig:spectrum}, can be found in \cite{Ayres:2026lyw,Ingo}. 
A storage cycle began with the UCN source proton-pulse and the filling of the storage vessel through the open beamport and experiment shutters. A short, $\SI{9}{\second}$ monitor interval, opening the butterfly shutter, was then recorded before the $\SI{180}{\second}$ storage period. Finally, the butterfly shutter was re-opened and the remaining UCNs were counted for $\SI{95}{\second}$, yielding on average 1.2 million counts per cycle. All time periods were optimized around the lowest statistical uncertainty, making a compromise with the allowed operation parameters of the UCN source.

In the data analysis, we performed 5$\sigma$-clipping of the counts after storage (normalized with the monitor counts) to reject cycles with outliers caused by proton beam interlocks, beam-current drops, UCN source control problems, or synchronization problems with the optical link for the detector readout. Data were also rejected in case the magnetic field control system was not able to compensate for external influence. 
The 5$\sigma$-clipping is justified by the assumption
that a genuine $n-n'$ signal would manifest in the form of a consistent magnetic-field-dependent drop in normalized counts, present in multiple
cycles and sequences at the same target field, and not as a single or sporadic undercount.


The measurements were performed within a time span of several months, during which the UCN source performance slightly drifted~\cite{Anghel2018,Doorenbos:2024phd}. During the data taking, the temperature of the vacuum tank was monitored by temperature sensors, which recorded both daily and seasonal fluctuations in the range of $\pm$\SI{5}{\celsius} (over the full period). 
We observed, on the one hand, a daily temperature modulation in the monitor-normalized counts. 
On the other hand, a close-to-linear temperature-correlated drift with a slope of~\SI{0.003}{\celsius}$^{-1}$, which latter was corrected via re-normalization, including error propagation. 
This drift could be attributed to changes with temperature of sizes of small gaps in the path of UCNs in the storage vessel and of the losses at wall reflections (see discussion in~\cite{Atchison:2007PhysRevC.76}), but also to the aforementioned variations in the UCN source output. All these affect the monitor and final counts slightly differently due to spectrum softening over time. The vacuum conditions were kept below 1$\times 10^{-5}$ mbar, at which additional losses due to residual gases were expected to be below the statistical uncertainty of the normalized counts per cycle~\cite{Ingo}.
To also minimize systematic errors from drifts in the UCN source yield and energy spectrum changes accompanying these, only data taken at most one day apart were combined into asymmetry values by pairing the ${i,j}$ field settings. These data groups led to a total of 832 magnetic-field setting comparisons, individually indicated by the indices ${i,j}$.

\begin{figure}
    \captionsetup{justification=Justified}

    \includegraphics[width=\linewidth]{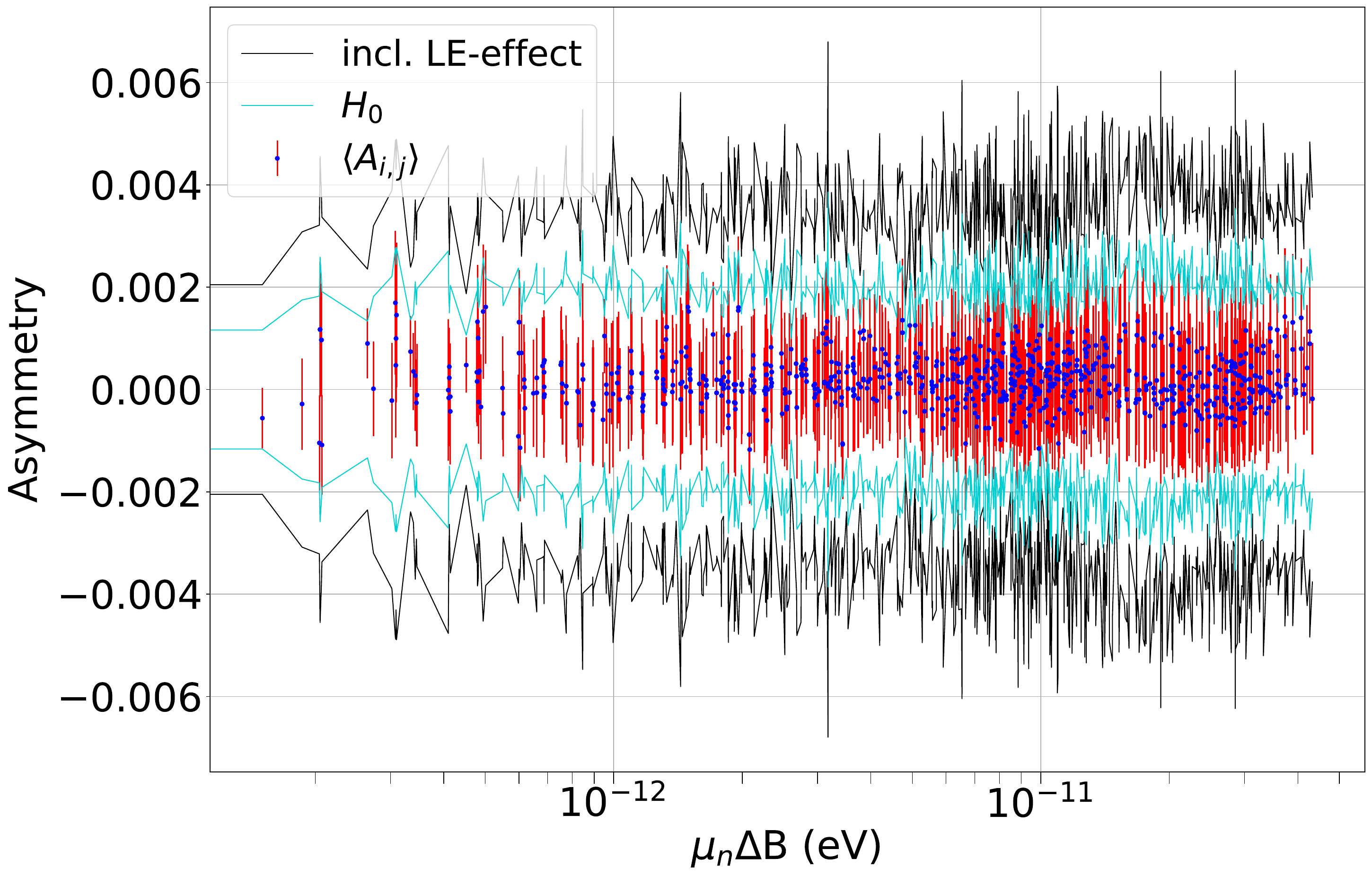}
    \caption{Mean asymmetry $\langle A_{i,j} \rangle$ (blue dots) and the standard errors (red bars) as a function of the magnetic field differences converted into energy unit. The cyan-blue line shows the limits of the null-hypothesis $H_0$ at 95\% C.L., the black outline shows the same limit after correcting for the look-elsewhere (LE) effect. $\langle A_{i,j} \rangle$ is consistent with $H_0$ within a standard error $\delta \langle A_{i,j} \rangle$.}
    \label{fig:asymmetry}
\end{figure}

The obtained asymmetries averaged over cycles $\langle A_{i,j} \rangle$  and the standard errors on the means $\delta \langle A_{i,j} \rangle$ are plotted in 
Fig.~\ref{fig:asymmetry}. The mean values are distributed within the boundaries of the null hypothesis, $H_0$. This is represented by the 95~\% C.L. of a Gaussian with a $\sigma$ equal to the measured uncertainty and centered at zero. After correcting for the look-elsewhere effect with the Bonferroni procedure~\cite{Dunn:1961} (considering 96 independent tests, one for each magnetic field
value), anomalous UCN losses were excluded within 95\% C.L.

\begin{figure*}

    \includegraphics[width=0.7\linewidth]{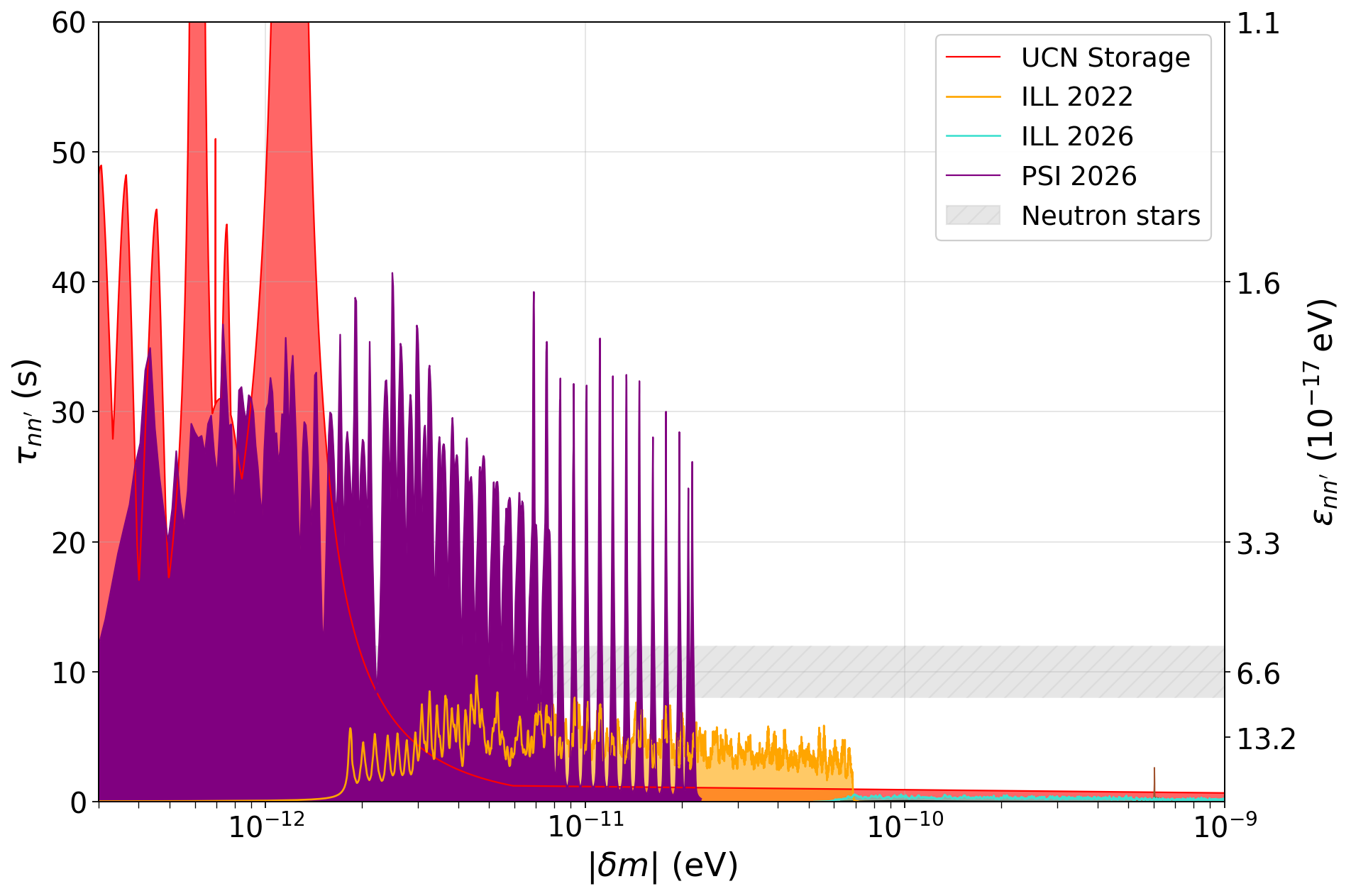}    \includegraphics[width=0.7\linewidth]{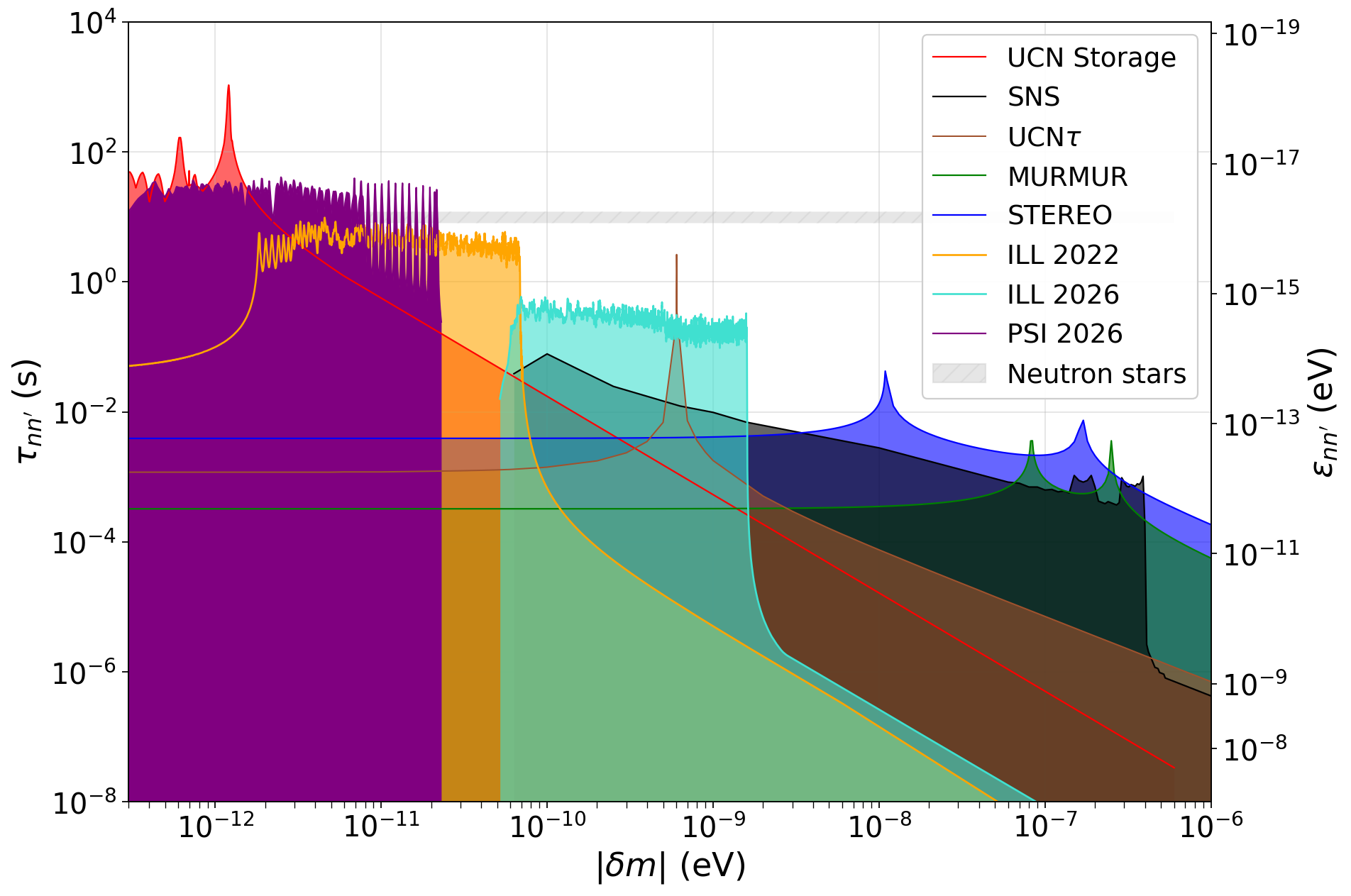}
    \captionsetup{justification=Justified}
    \caption{Top: Limits at 95\% C.L. on $\tau_{nn'}$ (linear axis) as a function of the mass splitting $\delta m$ based on the results of this analysis, \enquote{PSI-2026} (purple), along with UCN beam measurements at ILL (Grenoble) (yellow~\cite{Ban:2023cja} and cyan~\cite{BRENOT2026140673}), and the upper envelope of limits previously obtained from UCN storage measurements at PSI~\cite{nEDM:2020ekj} and ILL~\cite{Ban:2007tp,Serebrov:2008her,Bodek:2009zz,Berezhiani:2017jkn} 
    are shown in red. The astrophysical limit related to neutron star cooling \cite{Berezhiani:2020zck,McKeen:2021jbh,Berezhiani:2021src} is indicated by the gray shade.
    Bottom: Extended parameter ranges with additional limits from  \enquote{regeneration} experiments STEREO~\cite{Almazan:2021fvo} (blue), MURMUR~\cite{Stasser:2020jct} (green), SNS~\cite{Broussard:2021eyr,Gonzalez:2024dba} (black) and a reinterpretation of UCN$\tau$ data in~\cite{Hostert:2022ntu}. }
    \label{fig:limit_range}
\end{figure*}

We computed the $\tau_{nn'}$ limits based on Eq.~\eqref{eq:tau}  combining Monte Carlo simulations of $i,j$ resonance function pairs and the distributions for the asymmetry $i,j$ in the UCN counts. 
To obtain the latter, the asymmetry in Eq.~\eqref{eq:tau} was sampled as a Gaussian distribution centered at the measured  mean $\langle A_{i,j} \rangle$ with the standard error on this mean $\delta \langle A_{i,j} \rangle$ represented by the red error bars in  Fig.~\ref{fig:asymmetry}. For each of the 832 $B_{i,j}$ field pairs, the 95~\%~C.L. limit for $\tau_{nn'}$ was calculated as a function of the mass-difference $\delta m$.
In a next step, we took the upper envelope of all $\tau_{nn'}(\delta m)$ curves, i.e. used the best sensitivity per $\delta m$ value, as shown by the limit labeled \enquote{PSI-2026} in Fig.~\ref{fig:limit_range}.   
In the $\delta m$ range of 6 $-$ \SI{22}{\pico\electronvolt}, the sharp dips are caused by the widely spaced magnetic field settings, a consequence of a measurement-time compromise made to cover a larger interval and demonstrate the reach of sensitivity at higher $\delta m$ values as well.
Previous results are shown for comparison: UCN and cold neutron beam experiments~\cite{Broussard:2021eyr,Gonzalez:2024dba,Ban:2023cja} and reactor measurements~\cite{Almazan:2021fvo,Stasser:2020jct}, the latter ones also valid when $n$ and $n'$ states have substantial mass splitting. The astrophysical limit $\tau_{nn'} \sim 10$~s from constraints on the neutron star cooling \cite{Berezhiani:2020zck,McKeen:2021jbh,Berezhiani:2021src} is partly below the results obtained in the present work.
Along the horizontal axis of  Fig.~\ref{fig:limit_range}, a  possible contribution from a mirror magnetic potential is convertible as $\delta m/$peV =  $B'/\SI{16.58}{\micro \tesla}$.

To summarize, we have analyzed new and previous UCN-storage data, within a mass-splitting framework for $n-n'$ transitions. No magnetic-field-dependent loss signal was observed. The resulting 95\% C.L. limits on $\tau_{nn'}$ reach 20~s over parts of the $\delta m$ interval from 0.3 to \SI{22}{\pico\electronvolt}, extending the experimental coverage of the small mass-splitting region.


\medskip 

\textit{Acknowledgments} $-$
We thank M. Meier and L. Noorda for the excellent technical support, and are grateful for the valuable assistance of many support groups at PSI, especially the BSQ group operating the UCN source, the accelerator operating crews, the ‘Hallendienst’ and the mechanical workshops. 
We acknowledge the great support from the ETH IPA and D-PHYS mechanical workshops and the vocational training division. 

ETH and PSI appreciate the financial support from the Swiss National Science Foundation through projects 162574 (ETH), 169596 (PSI), 172626 (PSI), 172639 (ETH), 178951 (PSI), 188700 (PSI), 196416 (ETH), 200441 (ETH), 10003932 (ETH). 
We further acknowledge funding from the ETH Career Seed Grant SEED-13 20-2 and the SNF spark programme grant CRSK-2\_196416. 
The collaborators from the Jagiellonian University Cracow wish to acknowledge support from the National Science Center, Poland, under grant No. 2016/23/D/ST2/00715, No. 2018/30/M/ST2/00319 and No. 2020/37/B/ST2/02349, and also by the Minister of Education and Science under the agreement No. 2022/WK/07.
The work of Z.B. was supported in part by the research grant No.
2022E2J4RK ``PANTHEON: Perspectives in Astroparticle and
Neutrino THEory with Old and New messengers" under the program
PRIN 2022 funded by the Italian Ministero dell'Universit\`a e della
Ricerca (MUR) and by the European Union – Next Generation EU.

\end{document}